\documentclass[letterpaper,10pt]{article} 
\usepackage{osameet3} %% use version 3 for proper copyright statement
\usepackage{amsfonts,amssymb}
\usepackage{mdwmath,amsmath}
\usepackage[keeplastbox]{flushend}
\usepackage{cite}
\usepackage{mathtools}
\graphicspath{{../pdf/}{../jpeg/}}
\usepackage{color}
\usepackage{array}
\usepackage{enumitem}

\usepackage[utf8]{inputenc}
\usepackage{subcaption}

\usepackage{float}
\usepackage{wrapfig}

\usepackage[export]{adjustbox}

\usepackage{algorithmic}
\usepackage{relsize}

\renewcommand{\vec}[1]{\mathbf{#1}}
\let\oldhat\hat
\renewcommand{\hat}[1]{\oldhat{\mathbf{#1}}}

\usepackage{lipsum,booktabs,siunitx}
\newcolumntype{T}[1]{S[table-format=#1,group-digits=false]}
\allowdisplaybreaks

\newcommand{\ie}{\emph{i.e.}}
\newcommand{\eg}{\emph{e.g.}}
\usepackage{scalerel,stackengine}
\newcommand\equalhat{\mathrel{\stackon[1.5pt]{=}{\stretchto{%
    \scalerel*[\widthof{=}]{\wedge}{\rule{1ex}{3ex}}}{0.5ex}}}}

\usepackage[colorlinks=true,bookmarks=false,citecolor=blue,urlcolor=blue]{hyperref} %pdflatex

\begin{document}

\title{Low Complexity Convolutional Neural Networks for Equalization in Optical Fiber Transmission}

\author{Mohannad Abu-romoh$^{(1,\ast)}$, Nelson Costa$^{(2)}$, Antonio Napoli$^{(3)}$,\\ Jo\~ao Pedro$^{(2)}$, Yves Jaou\"en$^{(1)}$ and Mansoor Yousefi$^{(1)}$}
\address{$^{(1)}$ Telecom Paris, 19 Place Marguerite Perey, 91120 Palaiseau, France\\
$^{(2)}$Infinera Unipessoal Lda, Carnaxide, Portugal \; $^{(3)}$Infinera, London, United Kingdom}
\email{$^{(\ast)}$mohannad.aburomoh@telecom-paris.fr}

%% Uncomment the following line to override copyright year from the default current year.
\copyrightyear{2021}

\begin{abstract}
% 35 words
A convolutional neural network is proposed to mitigate fiber transmission effects, achieving a five-fold reduction in trainable parameters compared to alternative equalizers, and 3.5 dB improvement in MSE compared to DBP with comparable complexity.

\end{abstract}

\section{Introduction}
Distortions caused by the Kerr nonlinearity limit the achievable information rates (AIRs) at high powers in optical fiber communication. Mitigation of these distortions is possible using, \eg, digital back-propagation (DBP)~\cite{napoli2014reduced}. DBP, however, requires knowledge of fiber parameters and topology, and can be computationally expensive in part due to potentially large number of spatial segments. Artificial neural networks (ANNs) offer an alternative approach that might be less complex~\cite{8535376}. Learned digital backpropagation (LDBP) is proposed in~\cite{9240033}, in which the model is based on the split-step Fourier method (SSFM), and optimized using the standard learning algorithms for ANNs. In this paper, we consider a convolutional neural network (CNN) equalizer. We note that the SSFM coefficients are repeated in each span. Thus, instead of joint training of all neural network layers, we train a few unique layers that are shared in network depth. With this parameter sharing method, we substantially reduce the number of model trainable parameters. 
\vspace{-7pt}

\section{System Model}

\begin{figure}[H]
\vspace{-20pt}
\centering
\includegraphics[width=0.9\textwidth]{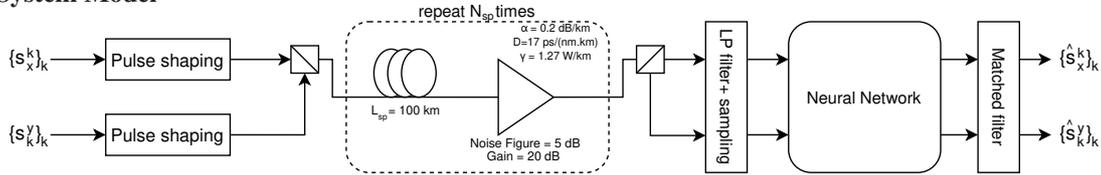}
\caption{\footnotesize Block diagram of the end-to-end system model}
\label{fig:system_model}
\vspace{-20pt}
\end{figure}

We assume the polarization division multiplexed (PDM) system shown in Fig.\ref{fig:system_model}. The signal carried by each polarization $U_{x,y}(t,z)$ is generated by quadrature amplitude modulation (QAM) $U_{x,y}(t,0) = \sum_{k=1}^{N_s} s_{x,y}^kp(t-kT_s)$, where $p(t)$ is the pulse shape as a function of time $t$, $\{s_{x,y}^k\}_k$ are transmitted symbols drawn from a constellation $\mathcal{S}$, $N_s$ is the number of symbols, and $T_s$ is the symbol period. The transmitted signal is propagated through an optical fiber modeled by the vector nonlinear Schr\"odinger's system 
$
    \frac{\partial U_{i}}{\partial z} = \big[-\frac{\alpha}{2} - \frac{\text{j}\beta_2}{2}\frac{\partial^2}{\partial t^2} + \text{j}\gamma\; (|U_{i}|^2+\frac{2}{3}|U_{\bar i}|^2)\big]\;U_{_{i}},
$
where $(i, \bar i)=(x,y)$ or $(y,x)$, $z$ is distance, and $\alpha$, $\beta_2$ and $\gamma$ are, respectively, loss, dispersion and nonlinearity coefficients.
\vspace{-8pt}

\section{Low Complexity Convolutional Neural Network Equalizer}
\begin{figure}[t]
\vspace{-40pt}
\centering
\includegraphics[width=\textwidth]{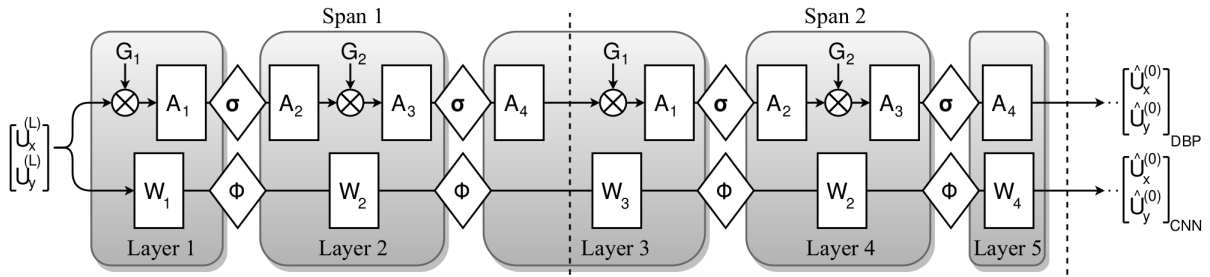}
\caption{\footnotesize An illustration of the proposed neural network architecture. The top branch corresponds to asymmetric DBP, and the bottom branch to the proposed CNN. $U_{x,y}^{(0)}$ and $U_{x,y}^{(L)}$ refer to the signals at the transmitting and receiving ends of the fiber, where $L$ is the number of segments.}
\label{fig:weight_sharing}
\vspace{-15pt}
\end{figure}

\vspace{-3pt}
In CNNs, each layer maps the input vector $\vec{X}^{(l-1)}$ to an output vector $\vec{X}^{(l)}=\Phi(\vec{\tilde W}^{(l)}*\vec{X}^{(l-1)})$ by convolution multiplication with a kernel $\vec{\tilde W}^{(l)}$. 
The main idea of this paper is to exploit the similarity between the neural network function and the un-folding of the SSFM, since both involve alternating between linear and nonlinear operations in their functions. It is hence possible to initialize the deep CNN parameters to perform SSFM; This approach has been studied and is referred to as model-based neural networks. We extend the model-based design of neural networks by assuming that some of the training parameters can be shared between multiple layers.
Fig.\ref{fig:weight_sharing} shows the proposed CNN equalizer with parameter sharing and the corresponding SSFM with $M=2$ steps per span. Since the linear and nonlinear operations of SSFM (\ie\; $A_m$ and $\sigma$ in Fig.\ref{fig:weight_sharing}, respectively) are repeated in each span in the optical fiber, we are able to identify 4 layers with unique parameters, namely $W_1\equalhat G_1A_1$, $W_2\equalhat A_2G_2A_3$, $W_3\equalhat A_4G_1A_1$ and $W_4\equalhat A_4$. Overall, the number of uniquely defined layers is $M+2$ regardless of the number of spans in the optical link. The nonlinear activation function used in our CNN is $\Phi(\vec{X}) = \vec{X}e^{\text{j}(\lvert\lvert \vec{X}\rvert\rvert^2+\frac{2}{3} \lvert\lvert \vec{Y}\rvert\rvert^2)}$ and $\Phi(\vec{Y}) = \vec{Y}e^{\text{j}(\lvert\lvert \vec{Y}\rvert\rvert^2+\frac{2}{3} \lvert\lvert \vec{X}\rvert\rvert^2)}$ for each polarization. In our model, amplification is applied after each segment such that $\prod_{k=1}^M G_k= 1$. With this, we provide the neural network the ability to adjust the intensity of the signal, and hence, the nonlinear phase rotation without the need of trainable parameters in the activation function.

\vspace{-4pt}
\section{Results}
\vspace{-2pt}
\begin{wrapfigure}[16]{r}[-10pt]{0.4\textwidth}
\vspace{-12pt}
\includegraphics[width=1\linewidth]{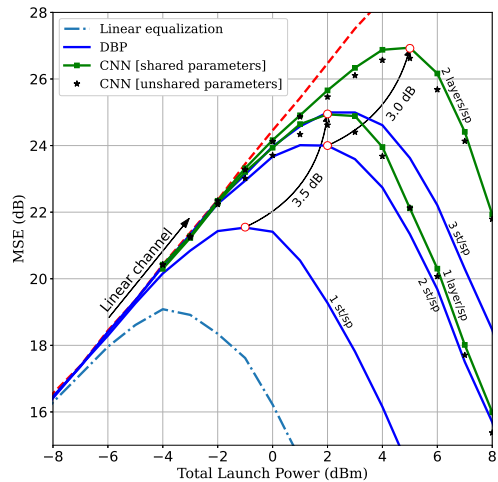}
\vspace{-21pt} 
\centering
\caption{\footnotesize{Comparison between compensation algorithms.}}
\label{fig:wrapfig}
\end{wrapfigure}
In our simulations, we generate dual-polarized 16-QAM signals at 32 Gbaud using root raised cosine pulse-shaping (roll-off factor = 0.1). We set the fiber length $L_f$=1000 km, and the number of spans $N_{sp}$=10. Forward-propagation is simulated with 50 steps/span, and data is initially sampled at a high sampling rate of 16 samples/symbol. On the receiver side, the signal is down-sampled to 2 samples/symbol before being processed by either CNN or DBP algorithms.
We use Keras API built on TensorFlow to implement and optimize the neural network parameters. During the training of the model, examples consisting of input-output pairs were randomly chosen from the set of launch powers $P\in\{$0,1,2,3,4$\}$ dBm. For each point in Fig.\ref{fig:wrapfig}, we generate 3300 input-output pairs for testing. We express the performance gain in terms of the normalized mean squared error (MSE) defined as $\frac{\rvert\rvert X-\hat{X} \rvert\rvert^2 + \rvert\rvert Y-\hat{Y} \rvert\rvert^2}{\rvert\rvert X \rvert\rvert^2+\rvert\rvert Y \rvert\rvert^2}$. The proposed CNN with 1 layer/span achieved MSE of $25$ dB at 2 dBm launch power which corresponds to $3.5$ dB improvement compared to the best performance attained by DBP with optimized step-size and equivalent complexity. For 2 layers/span, the peak MSE value for the CNN is $27$ dB measured at 5 dBm launch power, which is $3$ dB higher than DBP with similar complexity.
Comparing to LDBP (asterisks in Fig.\ref{fig:wrapfig}), a small gain in MSE is observed. We explain this gain by the neural networks ability to generalize better with fewer training parameters. In terms of training complexity, our model uses 3 uniquely defined layers comparing to 11 layers for LDBP, in the 1 layer/span setup, and 4 uniquely defined layers comparing to 21 layers for LDBP, in the 2 layers/span setup.

\vspace{-4pt}
\section{Conclusion}
\vspace{-4pt}

A parameter sharing method is proposed to reduce the training complexity of CNNs for equalization in optical fiber. The proposed approach yields 3 – 3.5 dB gain in MSE compared to optimized DBP with comparable complexity, and five-fold reduction in number of trainable parameters compared to LDBP at the same MSE.

\vspace{-6pt}
\section*{Acknowledgement}
\vspace{-3pt}
This project has received funding from the European Union’s Horizon 2020 research and innovation programme under the Marie Skłodowska-Curie grant agreement No. 813144.

\vspace{-4pt}
\footnotesize
\bibliographystyle{IEEEtran}
\bibliography{main}

\end{document}